\documentclass[pra,twocolumn,showpacs]{revtex4}
\usepackage{graphicx}
\date{\today}
\begin{document}
\title{Attosecond light pulses to reveal the time-dependent
rovibrational motion of the correlated electron pair in helium}

\author{Toru Morishita$^1$, Shinichi Watanabe$^1$ and C. D. Lin$^2$}
\affiliation{
$^1$Department of Applied Physics and Chemistry,
University of Electro-Communications, 1-5-1 Chofu-ga-oka,
Chofu-shi, Tokyo 182-8585, Japan\\
$^2$Department of Physics, Kansas State University, Manhattan, KS
66506,USA}
\begin{abstract}
We illustrate how attosecond light pulses can be used to directly mapping
out the time-dependence of the correlated motion of two excited
atomic electrons, discuss how the two-electron correlations manifest
themselves in realistic attosecond measurements, and propose the following
for experimental
exploration:
 (a) The single ionization signals which directly
reveal bending-vibrational motion of the correlated electron pair,
 (b) and also its rotational motion.
 (c) The double ionization signals which directly reveal
      the two-electron density in momentum space.
To facilitate the description of the above points, use is made of simple
wave packets of doubly-excited states of helium.
\end{abstract}

\pacs{32.80.Qk,32.80.Fb,31.25.Jf,31.15.Ja}

\maketitle
Following the recent developments in intense laser physics,
ultra-short XUV (extreme ultraviolet) light pulses with duration
of several hundred attoseconds (or a few atomic units) have been
reported\cite{hhg,krausz,shuntaro}. Such pulse durations are
comparable to the time scale of the electronic motion in the
ground and the lower excited states of atoms and molecules, thus
opening  up the route to the time-resolved study of electron
dynamics in matter, akin to the time-resolved tracking of the
atomic motion in a molecule enabled by the advent of femtosecond
laser pulses\cite{zewail}. In terms of probing electron dynamics,
the only time-domain measurement reported so far is the
determination of Auger lifetime by Drescher {\it et al.}\cite{krausz},
although other applications of attosecond pulses have been
reported elsewhere\cite{att2,att3}. Unlike molecules where the
time-dependent rotational and vibrational motions have clear
classical meaning, the significance of the time-dependence of the
electronic motion is presently elusive. The root of this misapprehension
lies
in the shell model of atoms which currently forms the basis for
interpretation of
almost all
the energy-domain measurements. According to this model, each
electron is moving in an effective central-field  potential made of the
electron-nucleus interaction plus an average potential due to the
remaining electrons.
Thus the time dependence of the interaction
of one electron with the others is often probed in the form of a
relaxation, which tends to be monotonic. Standard atomic structure
theory accounts for the deviations of the central field model in
terms of configuration interaction (CI). Thus in the time-domain
study of electron dynamics, the time-dependent CI coefficients are
calculated\cite{greek}. The point is that such coefficients
carry little physical meaning, and one must resort to some alternative
viewpoint for a physical interpretation.

In this Letter we suggest that a good place to study the
time-domain electron dynamics is where the breakdown of the shell model
is most severe, namely to probe multiply excited
states of an atom using attosecond XUV pulses. Theoretical
studies in the past decades have revealed that the shell model
fails completely for these states\cite{fano63,HKS,
ezra,Lin86,shin2e,toru223,Madsen,toru4e,Madsen2},
and the motions of electrons in these states are better
described by drawing analogy with the rotation and vibration of a floppy
polyatomic molecule, where the periods are of the order of a few to sub-
femtoseconds. By using doubly excited states of helium as an
example, we show that the rotational and bending vibrational
motions of the two electrons can be probed directly with XUV
attosecond pulses, where
single and/or double ionization signals of coherently populated
doubly excited states should reveal these rotational and/or
vibrational motions directly. For these systems, the ``movies'' of
the correlated motion of the
two electrons  can be made
to reveal their time evolution, similar to the ``movies'' showing the
motion of atoms in a molecule. However, there is a significant
difference. The
rotational and the vibrational periods in a molecule are
at least two orders of magnitude different. Thus their motions
are not directly coupled. For atoms, the rotational and
vibrational periods are comparable. To disentangle the various
rotational and bending vibrational modes would require a careful
preparation of the initial coherent double excited states. Addressing
  the preparation of such coherent doubly excited
states is not a purpose of this Letter. Instead, we consider
simplest initial coherent states and calculate the time-dependence
of the ionization yield to show that they indeed reveal the
rotational and the bending motion of the two electrons directly.

To describe the collective motion of the two electrons in the doubly
excited
states\cite{Lin86}, it is
convenient to look at them as a linear XY$_2$ triatomic
molecule, with X playing the role of the nucleus and Y an electron. Instead
of the independent electron coordinates ${\bf r}_1$ and ${\bf r}_2$,
the wavefunctions are to be expressed in terms of $R$,
$\Omega_v=(\alpha,\theta_{12})$ for the internal stretching and bending
vibrational  motions,
and three Euler angles
$\Omega_r=(\alpha^\prime,\beta^\prime,\gamma^\prime)$ for the
overall rotational  motions. Here the hyperradius $R$ and the
hyperangle $\alpha$  are defined by $r_1=R\cos\alpha$ and
$r_2=R\sin\alpha$. Thus $R$ stands for the size of the atom,
$\alpha$ measures the relative distances of the two electrons from
the nucleus, and $\theta_{12}$ is the angle which the two electrons make
with the nucleus at the vertex.  Note that each
wavefunction depends on six variables excluding spins. For visualization,
the
vibrational and rotational mode of each state can be defined via
the densities,
\begin{eqnarray*}
\rho^{\rm vib}_j(\alpha,\theta_{12})&=&\int|\varphi_j|^2 d\Omega_r dR,\\
\rho^{\rm rot}_j(\Omega_r)&=&\int|\varphi_j|^2 d\Omega_v dR
\end{eqnarray*}
for each state $\varphi_j$. In presenting $\rho^{\rm rot}(\Omega_r)$,
we choose $\hat {\bf r}_1-\hat {\bf r}_2$ and
$\hat {\bf r}_1\times\hat {\bf r}_2$ to be parallel to the
$z'$- and $y'$-axes of the body-fixed frame, respectively.
We consider linearly polarized laser pulses parallel to the $z$-axis
of the space-fixed frame. Thus, we only need to consider
the magnetic component of the total angular momentum, $M=0$,
so that the density does not depend on $\gamma^\prime$.

\begin{figure}[ht]
\includegraphics[scale=0.38]{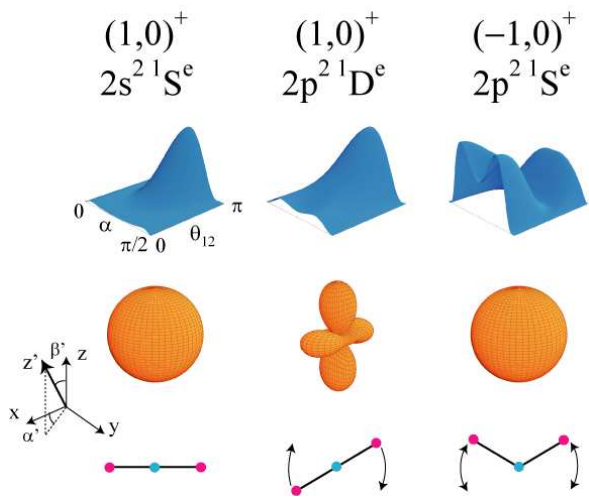}
\caption{(color online). Density plots of the doubly excited states. The top row
presents the relief plots of  $\rho^{\rm vib}(\alpha,\theta_{12})$, and
the middle row
shows $\rho^{\rm rot}(\alpha^\prime,\beta^\prime)$ by polar plots. As seen
 from the
density plots, $2s^2~^1S^e$, $2s^2~^1D^e$, and  $2p^2~^1S^e$
are considered as ground, rotationally excited, and
vibrationally excited states, respectively. The bottom row
sketches the corresponding classical rotational and bending vibrational
modes of
the two correlated electrons.} \label{fig:basic}
\end{figure}

Neglecting autoionization for the time being, consider first the
``stationary'' doubly excited states, conventionally called
$2s^2~^1S^e$, $2p^2~^1S^e$, and $2p^2~^1D^e$ states. These states
exhibit large configuration mixing and are more accurately
classified using the $(K,T)^A$ quantum numbers as described in
\cite{Lin86}. In this designation, the three states are denoted as
$(K,T)^A=(1,0)^+$, $(-1,0)^+$, and $(1,0)^+$, respectively. In
Fig.~\ref{fig:basic} the vibrational and rotational densities of
each state are given, together with a classical sketch of the
motion of the two electrons.

For $2s^2~^1S^e$, $\rho^{\rm vib}(\alpha,\theta_{12})$ has a
maximum at $\theta_{12}=\pi$ and $\rho^{\rm rot}(\Omega_v)$ is
isotropic. Thus, this state is considered as the ground state of
the rovibrational motion. For $2p^2~^1D^e$, $\rho^{\rm
vib}(\alpha,\theta_{12})$ has maximum at $\theta_{12}=\pi$ similar
to $2s^2~^1S^e$, but $\rho^{\rm rot}(\Omega_r)$ has nodes in
$\beta^\prime$ similar to $|Y_{20}(\beta^\prime,\alpha^\prime)|^2$
due to total angular momentum $L=2$ with $T=0$. That is, this
state is rotationally excited. For $2p^2~^1S^e$, on the contrary,
$\rho ^{\rm rot}$ is isotropic, but there is a nodal line in
$\rho^{\rm vib}(\alpha,\theta_{12})$ at about $\theta_{12}=\pi/2$,
so that this state is an excited state in the bending-vibrational
mode. With this understanding, we proceed to discuss how to
observe the time-dependence of these rotational and vibrational
motions by first creating coherent states by a pump pulse. We will
then consider the use of attosecond light pulses for probing the
time dependence of the coherent state via the ionization yield. To
this end, we calculated the ionization probability from such
coherent doubly excited states by applying the time-dependent
Hyperspherical method\cite{TDHS} to the single ionization, and the
first order perturbation theory to the double ionization
separately.

\begin{figure}[ht]
\includegraphics[scale=0.28]{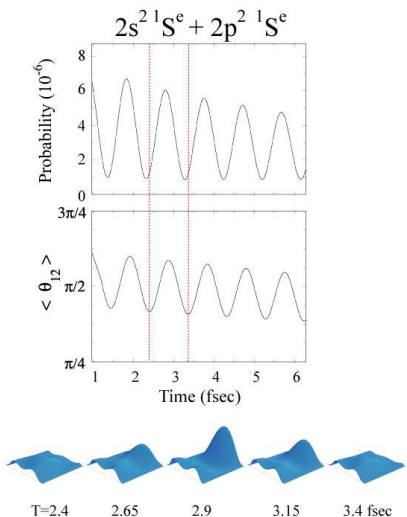}
\caption{(color online). Delay time dependence of (a) stimulated single ionization
probability from a coherent doubly excited state of
$\left[\varphi(2s^2~^1S^e)+2\varphi(2p^2~^1S^e)\right]/\sqrt{5}$ of He,
(b) the average angle between the two electrons with respect to
the nucleus, and (c)the two electron density distributions over one
period.}
\label{probe_t12}
\end{figure}

We highlight the following three points in turn. {\it (1)
Attosecond pulses probing bending vibrational motion.} When the
$2s^2~^1S^e$ and $2p^2~^1S^e$ states are coherently excited at
$t=0$, a vibrational wave packet is created. We used a Gaussian
attosecond pulse of mean energy of 21.8 eV, duration of 142 asec,
and intensity of $3.5\times 10^{12}$ W/cm$^2$   to ionize this
coherent state by stimulated emission. The resulting ionization
yield of He$^+(1s)$+e in the energy region of $-2$ to $-1$ a.u.,
as shown in the top frame of Fig.~\ref{probe_t12}, shows the
expected oscillations as a function of delay time. The oscillation
has a period of about 970 asec, corresponding to the inverse of
the energy separation between the two states, $2\pi/\Delta E$. In
fact, this oscillation can be traced directly to the bending
vibrational motion, as seen clearly from the calculated time
evolution of the average angle between the two electrons with the
nucleus at the vertex, $\langle\theta_{12}\rangle$. At the bottom
of this figure, the calculated $\rho^{\rm
vib}(\alpha,\theta_{12})$ is shown over a period. The ionization
probability peaks when the two electrons are mostly on opposite
sides of the nucleus. We note that the ionization yield decreases
with increasing delay time, due to the autoionization of the
$2s^2~^1S^e$ state which has a lifetime of about 5.4 fsec. Thus
the ionization yield can be used to measure the lifetime of the
shorter-lived state directly without the help of an IR
laser\cite{krausz}. Note that attosecond pulses are needed in
order to see the identifiably rapid oscillations in the ionization
probabilities.

\begin{figure}[ht]
\includegraphics[scale=0.28]{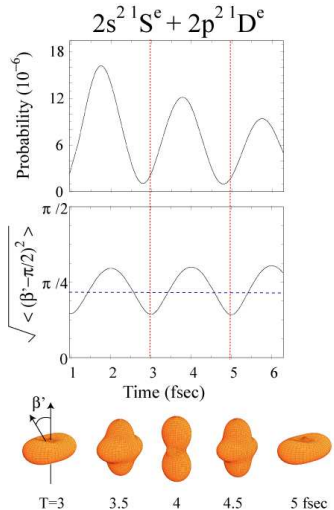}
\caption{(color online). Delay time dependence of (a) stimulated ionization
probability from a coherent doubly excited state of
$\left[\varphi(2s^2~^1S^e)+\varphi(2p^2~^1D^e)\right]/\sqrt{2}$ of He,
(b) the average of the angle between the
two-electron axis with respect to the laser polarization (see text),
and (c) the rotational distributions of the atom in Euler angles.}
\label{probe_s1d1}
\end{figure}

{\it (2) Attosecond pulses probing the rotational motion.}
In Fig.~\ref{probe_s1d1}  we show the ionization probability
of He$^+(1s)$+e in the energy region of $-2$ to $-1$ a.u.
vs delay time from a coherent state made of $2s^2~^1S^e$
and $2p^2~^1D^e$. In this case each state is the ground state of the
vibrational motion. The time-dependent oscillation is to be traced to the
rotational motion, since the  $2p^2~^1D^e$ state  has total angular
momentum
$L=2$ with $T=0$, {\it i.e.}, it is rotationally excited in $\beta^\prime$.
The time dependence of the ionization probability is shown to follow
the oscillation of the Euler angle $\beta^\prime$.
(The average of the deviation from $\pi/2$.)
This angle $\beta^\prime$ is defined to be the
angle between the ``molecular axis'' (the axis between the two electrons)
with respect to the laser polarization direction. In this case,  more
ionization occurs when the line joining the two electrons are
parallel to the laser polarization. A more complete view of the
rotational motion of the wave packet can be seen from the bottom
frame where the rotational density evolution within one period is
displayed. Note that the oscillation period (2.0 fsec) is inversely
proportional to the energy separation between the two states.

{\it (3)Attosecond pulses probing two-electron wavefunctions in
momentum space.}
The oscillation of the single ionization probabilities presented
above is generic to any two-state system. The oscillation
can be interpreted in terms of bending vibrational or
rotational motions of the two electrons only
when they are accompanied by such careful theoretical analysis as
presented here.
Without {\it a priori} information about the nature of the coherent
state, a blind display of the six-dimensional two-electron
wavefunction would not be able to reveal the nature of  the two-electron
dynamics.
  For an arbitrarily created coherent
state, an ambiguous analysis of this sort would be
rather difficult. We thus ask whether it is possible to map out
the time-dependence of the two-electron motion
directly. One method we propose is to measure double ionization of
helium by an attosecond light pulse.

\begin{figure}[ht]
\includegraphics[scale=0.48]{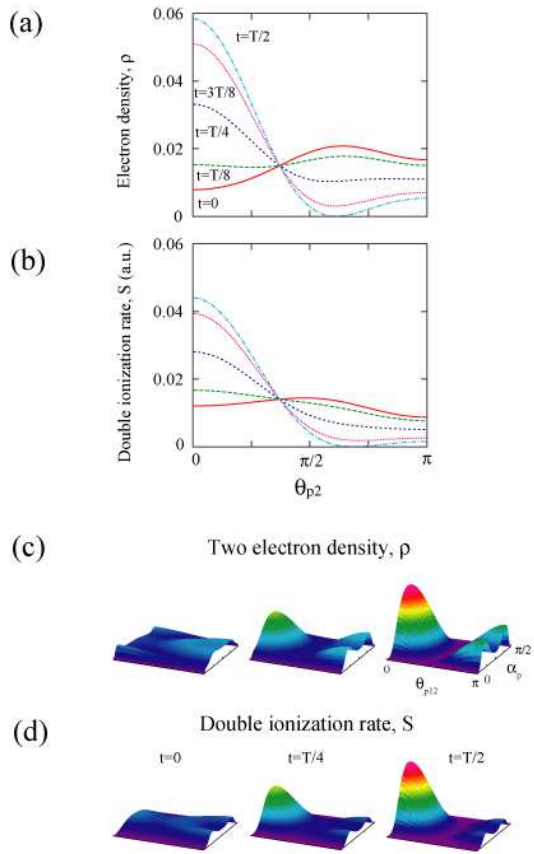}
\caption{(color online). Time-dependence of double ionization from a coherent
state of $\left[\varphi(2s^2~^1 S^e)+\varphi(2p^2~^1
S^e)\right]/\sqrt{2}$. (a) The angular distribution of the
electron density of the initial coherent state in momentum space.
Shown is the momentum density of the second electron with respect
to the direction of the fixed first electron; (b) The distribution
of the calculated ionization signal of the second electron with
respect to the direction of the fixed first electron; (c) The
two-dimensional two-electron momentum space density distributions
of the initial coherent state; (d) Double ionization signals
showing the relative momentum distributions between the two
ionized electrons.  (see text).} \label{double_ionization}
\end{figure}

We calculated the probability of double ionization by an attosecond
light pulse with mean energy of 27.2 eV and pulse duration of
83 asec. We use a weak light pulse so that double ionization
probability is calculated by the first order perturbation
theory. In the actual calculation, the velocity gauge was used
and the electric field of the attosecond pulse was assumed Gaussian.
The two free electrons in the final states are approximated by
properly symmetrized products of plane waves. Under this
approximation, one can formally show that in the limit of a very narrow
pulse,
the ionization probability is proportional to the
two-electron charge density in the momentum space of the initial
coherent state.

We have carried out a numerical calculation for the simplest coherent state
made of  $2s^2~^1 S^e$ and $2p^2~^1S^e$. The double ionization
probability $P({\bf p}_1,{\bf p}_2,t)$
is a function of the six-dimensional  momenta of the two electrons.
Here $t$ is the time that elapsed after the pump pulse.
This particular example of a coherent state represents a bending
vibrational wave packet.
In Fig.~\ref{double_ionization}(a) we plot the ionization rate
$S({\bf p}_1,{\bf  p}_2,t)\propto P({\bf p}_1,{\bf  p}_2,t)/
\left[\hat\epsilon\cdot({\bf p}_1+{\bf p}_2)\right]^2$ as a function of
$\theta_{p2}$, with fixed ejected electron energies of two electrons
$e_1=e_2=2.2$ eV and $\theta_{p1}=0$, over half a period
at $t=0$, $T/8$, $T/4$, $3T/8$, and $T/2$ asec (the total double
ionization probability is expected
to oscillate with a period of $T=970$ asec for this coherent state)
 Here, the angles are
measured relative to the laser polarization $\hat\epsilon$. In the
upper frame, the density plots constructed from the momentum space
wavefunction are shown. It is clear that the angular dependence of
the ionization signal resembles the angular dependence of the
momentum space electron density; see
Fig.~\ref{double_ionization}(b). Note that the amplitude of the
oscillations in the ionization signal is somewhat weaker. This is
due to the averaging effect from the finite pulse duration of 83
asec. (The mean energy of 27.2 eV was used, since such short
attosecond pulses have to be generated from the plateau region of
the high-order harmonics.)

We define the
momentum space hyperspherical coordinates for the two
electrons which enable us to similarly define the momentum space
counterpart of the vibrational
and rotational densities.
In Fig.~\ref{double_ionization}(c) and (d) we compare
the ionization measurement with the momentum space densities
as functions of $\alpha_p=\arctan(p_2/p_1)$ and
$\theta_{p12}=\arccos (\hat{\bf p}_1\cdot\hat{\bf
p}_2)$\cite{kato,peterkop}.
(Note that these angles are not conjugate to the
hyperspherical coordinates variables in the configuration space,
$\alpha$ and
$\theta_{12}$.) Indeed, the ionization
signal  clearly reproduces the bending vibrational density of the
two electrons in the momentum space.
Representing the final state wavefunctions by
plane waves as in the present example is undoubtedly an
over-simplification.
Improved calculations might show slight distortions to the double
ionization signals, a task left for future exploration.
Nevertheless, it is undeniable that the present calculation points
out that attosecond pulses are capable of mapping out the two-electron
dynamics. Such information cannot be directly derived
 from energy-domain measurements.

In summary, we have shown that attosecond light pulses can be
used to probe the correlated motion of two excited electrons. For
properly created coherent states the rotational and/or bending
vibrational modes of the two excited electrons can be directly
mapped by the single or double ionization signals vs the time
delay. Extension of the present analysis to multiply excited
states\cite{toru223,toru4e} and many-body systems would similarly offer a
new perspective that distinguishes itself from the established
energy-domain measurements.

This work was supported in part by a Grant-in-Aid for Scientific
Research (C) from the Ministry of Education, Culture, Sports,
Science and Technology, Japan, and by the 21st Century COE program
on ``Coherent Optical Science''.
This work was also supported by Chemical Sciences, Geosciences and
Biosciences Division, Office of Basic Energy Sciences,
Office of Science, U.S. Department of Energy.
TM was also supported by financial aids from the Matsuo Foundation
and the University of Electro-Communications.
TM thanks Prof. H. Kono, Dr. X. M. Tong and Dr. Z.~X.~Zhao for useful
discussions.


\begin{references}
\bibitem{hhg}
M. Hentschel {\it et al.}, Nature (London) {\bf 414}, 509 (2001).
\bibitem{krausz}
M. Drescher {\it et al.}, Nature (London) {\bf 419}, 803 (2002).
\bibitem{shuntaro}
T. Sekikawa, {\it et al.} Nature (London) {\bf 432}, 605 (2004).
\bibitem{zewail}A. H. Zewail, J. Phys. Chem. {\bf 104}, 5660
(2000).
\bibitem{att2} E. Gouliemakis {\it et al.}, Science {\bf 305}, 1267
(2004).
\bibitem{att3} F. Lindner {\it et al.}, Phys. Rev. Lett. {\bf 95},
040401 (2005).
\bibitem{greek} C. E. Nicolaides, {\it et al.}, Phys. B{\bf 35}, L271 (2002)
\bibitem{fano63}
J.~W.~Cooper, U.~Fano, and F.~Prats, Phys. Rev. Lett. {\bf 10}, 518 (1963).
\bibitem{HKS}
D.~R.~Herrick and O.~Sinano\u{g}lu, Phys. Rev. {\bf 11}, 97 (1975);
M.~E.~Kellman and D.~R.~Herrick, J. Phys. B {\bf 11}, L755 (1978);
O.~Sinano\u{g}lu and D.~R.~Herrick, J. Chem. Phys. {\bf 62}, 886 (1975);
M.~E.~Kellman and D.~R.~Herrick, Phys. Rev. A {\bf 22}, 1536 (1980).
\bibitem{ezra}
G.~S.~Ezra and R.~S.~Berry, Phys. Rev. A {\bf 28}, 1974 (1983).
\bibitem{Lin86}
C. D. Lin, Adv. At. Mol. Phys. {\bf 22}, 27 (1986).
\bibitem{shin2e}
S.~Watanabe and C.~D.~Lin, Phys. Rev. A {\bf 34}, 823 (1986).
\bibitem{toru223} T. Morishita and C. D. Lin, Phys. Rev. A {\bf 67},
022511 (2003).
\bibitem{Madsen}
L. B. Madsen, J. Phys. B {\bf 36}, R223 (2003).
\bibitem{Madsen2}
M. D. Poulsen and L. B. Madsen, Phys. Rev. A {\bf 71}, 062502 (2005).
\bibitem{toru4e} T. Morishita and C. D. Lin,
Phys. Rev. A {\bf 71}, 012504 (2005).
\bibitem{TDHS}
T. Morishita, K. Hino, T. Edamura, D. Kato, S. Watanabe and M. Matsuzawa
J. Phys. B: At. Mol. Opt. Phys. {\bf 34}, L475 (2001).
\bibitem{kato}
D. Kato and S. Watanabe, Phys. Rev. A {\bf 56}, 003687 (1997).
\bibitem{peterkop}
R. Peterkop, Theory of Ionization of Atom by Electron Impact (Colorado
Associated University Press, Boulder, 1977).
\end{references}
\end{document}